\documentclass{aims}
\usepackage{amsmath,mathtools,amsthm,mathrsfs}
\usepackage{subcaption}
  \usepackage{paralist}
  \usepackage{graphics} 
  \usepackage{epsfig} 
\usepackage{graphicx}  \usepackage{epstopdf}
 \usepackage[colorlinks=true]{hyperref}
\hypersetup{urlcolor=blue, citecolor=red}

\def\currentvolume{X}
 \def\currentissue{X}
  \def\currentyear{200X}
   \def\currentmonth{XX}
    \def\ppages{X--XX}

\theoremstyle{plain}
\newtheorem{theorem}{Theorem}[section]
\newtheorem{lemma}[theorem]{Lemma}
\newtheorem{corollary}{Corollary}[theorem]
\newtheorem{claim}{Claim}[section]
\theoremstyle{definition}
\newtheorem{definition}{Definition}[section]

\newtheorem{example}{Example}[section]
\newtheorem{assumption}{Assumption}[section]
\newtheorem{remark}{Remark}[section]

\usepackage{graphicx}
\usepackage{lmodern}
\usepackage{amssymb}
\usepackage{pst-node}
\usepackage{tikz-cd}

\newcommand{\B}{\mathscr{B}}

\newcommand{\E}{\mathbb{E}}
\newcommand{\g}{\mathcal{G}}
\newcommand{\I}{\mathscr{I}}
\newcommand{\id}{I}
\newcommand{\K}{\mathcal{K}}
\renewcommand{\l}{\bar{l}}
\renewcommand{\L}{\mathcal{L}}
\newcommand{\lb}{\bar{\mathcal{L}}}
\renewcommand{\O}{\mathcal{O}}
\newcommand{\Phig}[1]{\phi_{#1}}
\newcommand{\T}{P}
\renewcommand{\v}{\bar{v}}
\newcommand{\Y}{\mathcal{Y}}

\DeclareMathOperator{\hor}{\text{hor}}
\DeclareMathOperator{\Hor}{\text{Hor}}
\DeclareMathOperator{\ver}{\text{ver}}
\DeclareMathOperator{\Ver}{\text{Ver}}

\renewcommand{\triangleq}{\coloneqq}

\newcommand{\R}{\mathbb{R}}

\def\SOthree{\mathop{\mathbb{SO}(3)}}
\def\sothree{\mathfrak{so(3)}}

\def\SEthree{\mathop{\mathbb{SE}(3)}}
\def\Sone{\mathop{\mathbb S^1}}

\newcommand{\set}[2]{ \left\{ #1 : #2 \right\} }
\def\tr{\mathop{\hbox{tr}}}
\newcommand{\euc}[1]{\bar{#1}}
\def\ra{{}\rightarrow{}}

\let\oldrho\rho
\renewcommand{\rho}{\tau}

\renewcommand{\varphi}{\oldrho}

\let\oldhat\hat
\renewcommand{\hat}[1]{{#1}^{\wedge}}
\renewcommand{\widehat}[1]{{#1}^{\wedge}}

\let\oldtilde\tilde
\renewcommand{\tilde}[1]{\oldhat{#1}}
\renewcommand{\widetilde}[1]{\oldhat{#1}}

\title[Exploiting Symmetry in Observer Design] 
      {A bundle framework for observer design on smooth manifolds with symmetry}

\author[A. A. Joshi, D.H.S. Maithripala and R. N. Banavar]{}


 \email{anantjoshi@iitb.ac.in}
 \email{sanjeevam@sltc.ac.lk}
 \email{banavar@iitb.ac.in}

\thanks{The authors would like to thank the Indian Institute of Technology Bombay and the Sri Lanka Technological Campus,  Padukka, for their support both logistical and financial.}

\thanks{$^*$ Corresponding author}

\begin{document}
\maketitle

\centerline{\scshape Anant A. Joshi}
\medskip
{\footnotesize
 \centerline{Department of Mechanical Engineering,}
   \centerline{Indian Institute of Technology Bombay, Mumbai}
   \centerline{ Maharashtra, 400069,  India}
} 

\medskip

\centerline{\scshape D.H.S. Maithripala}
\medskip
{\footnotesize
 \centerline{ Department of Mechanical Engineering,}
   \centerline{University of Peradeniya,}
   \centerline{KY20400, Sri Lanka.}
   \centerline{School of Postgraduate Studies,}
 \centerline{ Sri Lanka Technological Campus,  Padukka,}
   \centerline{CO 10500, Sri Lanka}
   
}

\medskip

\centerline{\scshape Ravi N. Banavar$^*$}
\medskip
{\footnotesize
 \centerline{Department of Systems and Control Engineering,}
   \centerline{Indian Institute of Technology Bombay, Mumbai}
   \centerline{ Maharashtra, 400069,  India}
} 

\bigskip

 \centerline{(Communicated by the associate editor name)}

\begin{abstract}
The article presents a bundle framework for nonlinear observer design on a manifold with a a Lie group action. 
	The group action on the manifold decomposes the manifold to a quotient structure and
	an orbit space, and the problem of observer design for the entire system gets decomposed to a design
	over the orbit (the group space) and a design over the quotient space. The emphasis throughout the article is on 
	presenting an overarching geometric structure; the special case when the group action is free is given special emphasis. Gradient based observer design on a Lie group  is given explicit attention. The concepts developed are illustrated by applying them on well known examples, which include the action of $\SOthree$ on $\R^3 \setminus \{0\}$ and the simultaneous localisation and mapping (SLAM) problem. 		
\end{abstract}

\section{Introduction}
	Observer design and estimator design have enjoyed a long history after the appearance of seminal work \cite{kalman-1960,kalman-1961,luenberger-1964}. 	
	The Kalman filter was developed for linear systems but it has been modified and applied to various other systems as well. Much of estimation theory in engineering has centred around the Kalman filter. The setting of the problem is a vector space and the tools involve linear systems theory. 
	Nonlinear extensions to the Kalman filter have not pushed the domain of theoretical ideas, but have largely been restricted to ideas like linearization, such as the, extended Kalman filter (EKF) \cite{junkins-optest}, unscented Kalman filter \cite{junkins-optest,uhlmann-2004,uhlmann-1995}, and the multiplicative extended Kalman filter \cite{lefferts-1982,markley-2003,markley-fund}. The work \cite{crassidis-2007} presents more such techniques for attitude estimation.
	The dynamics of a large class of mechanical and aerospace
	systems, however, evolve in a nonlinear setting, and in particular, on smooth 
	manifolds or specifically on Lie groups. Control synthesis for such systems in an intrinsic 
	framework, that respects the geometry of the underlying manifold, has been 
	much studied in the past two decades \cite{bloch-2001,bloch-2000,bloch,bullo}. The parallel, observer design or
	estimator design, has received less attention. Our work focusses on the latter.
	
	There is a large body of work in non-linear observer design, see \cite{bonnabel-2011-fl,mahony-2015,sanner-2003,cunha-2007}.  
	Many aero-mechanical systems are mathematically modelled as systems evolving on 
	Lie groups. Intrinsic observers directly on the Lie group are designed to avoid the pitfalls of parametrization, like Euler angles (that suffers from singularity at particular 
	configurations), or quaternions (that suffer from over-parametrization of
	the rotation group).  
	The concept of fusing two measurements with different frequency characteristics to design an observer for attitude estimation directly on $\SOthree$ rather than using quaternions or Euler angles was presented in \cite{mahony-2008}. This complementary filtering was extended to the case when observations are made from the inertial frame and relayed to the agent in \cite{mahony-2009}, and the work also shows when observability can be achieved with a single direction measurement. Complementary filtering on the Lie groups $\mathbb{SE}(3)$ and $\mathbb{SL}(3)$ has also been done \cite{mahony-2007,mahony-2012}. The work of \cite{mahony-2009-icra} contains an observer on $\mathbb{SE}(3)$, that is similar in structure to what is presented in \cite{mahony-2008}.	
	 	
	The work reported in \cite{bonnabel-2011} demonstrates a separation principle on Lie groups for linearized controller and observer design. The invariant extended Kalman filter in \cite{bonnabel-2018,bonnabel-2007} applies the ideas of the conventional extended Kalman Filter to Lie groups. However, it also involves linearization of state error to propagate covariance (just as in a conventional EKF). {It has been studied further in \cite{bonnabel-2009-cdc}, in the setting of symmetry preserving observers presented in \cite{bonnabel-2008}.} Its stability properties have also been studied in \cite{bonnabel-2017,bonnabel-2017-ifac}. Extending these notions to the discrete setting, an intrinsic discrete EKF for Lie groups is presented in \cite{bourmaud-2013}. Motivated by the Kalman filter for linear systems, an estimator design for discrete time systems evolving on Lie groups perturbed by stochastic noise is also developed in \cite{bonnabel-2015}. 
	
	Gradient based techniques for designing observers are very intuitive to understand, since they can be visualised as a gradient descent kind of algorithm 
	moving to make the observer error zero. The work \cite{mahony-2010} introduces the idea of gradient based observer design for kinematic systems evolving on Lie groups assuming full state and input measurement while \cite{mahony-2017-icra} gives a gradient based observer for discrete time observer design on $\SOthree$. These ideas are also applied in \cite{mahony-2011} to the design of gradient based observer on $\mathbb{SE}(3)$ assuming velocity measurements and measurement of position of $n$ points whose inertial locations are known. Extending these results gradient based observers for simultaneous localisation and mapping (SLAM) have been developed in \cite{wang-2018,forbes-2018}. 	
	
 Similar to optimal control, optimal observer design has also received some attention. Near optimal (deterministic) filters (filters on which we know how far they are from optimality) have been developed for systems evolving on $\Sone,\mathbb{SO}(n)$ \cite{mahony-2009-opt,mahony-2011-tac}.  The work presented in \cite{mahony-2013-tac} considers optimal attitude filtering considering only kinematics and \cite{mahony-2015-tac} extends that to dynamics as well (on the tangent bundle of the Lie group).	 
 
While the 
manifold framework to observer design with group symmetry has been 
presented by quite a few individuals, all these contributions have been cited either before or 
after this paragraph, the overarching mathematical framework replete with all the tools - bundle theory, section assignment, connection,
vertical and horizontal spaces, flows on the base space and fibre space - 
has been distinctly missing. This article fills in this gap.
To encompass a large class of engineering systems of
interest, we focus on the case when there is a Lie group $G$ acting on the configuration manifold $P$ of the system and the system is  also invariant under this action. It is also assumed that the same Lie group acts on the output manifold so as to have a group action equivariant output.
 
 We briefly explain the mathematical machinery now. When a Lie group $G$ acts properly on a smooth manifold $P$, it provides a stratification of the  manifold where each stratum corresponds to an orbit, $\mathcal{O}(\cdot)$, of the group action. 
	The collection of these orbits, the modulo space of the equivalent classes of orbits $P/G$, is called the base space and has the structure of a smooth manifold when the action is free. 
	The triple $(P,P/G,\pi_\phi)$, where $\pi_\phi : P\to P/G$ is the canonical projection, has the structure of a principal bundle with structure group $G$ when the action is free (for free actions, a cross-section of this bundle allows one to associate with each point, $p\in P$, on the manifold a unique pair $([p],g)$ where $[p]\in P/G$ is called the base coordinate and $g\in G$ is called the fibre coordinate). When the action is not free, the problem becomes much more involved, but similar ideas can be developed with careful treatment.
An invariant vectorfield on $P$ induces a well defined unique vectorfield on the base space $P/G$. Thus the flow of such a vectorfield will carry orbits to orbits. Thus the flow an invariant vectorfield can be projected on to the base space $P/G$ and the group $G$. The projection onto $G$ depends on the cross section in a unique way. This splitting of the flow induces a corresponding splitting of the system into one that is evolving on the Lie group and one that is evolving on the  base space. Theorem \ref{thm:ReducedSystem} forms the crux of this decomposition
and presents two vector fields on equivalence classes - one in the quotient space (base space) arising from the
bundle structure, and the other due to the group action at any give point 
on the manifold. This feature has not been illustrated in any existing literature so
far.

	The mathematical structure presented in the previous paragraph can naturally be applied to 
	observer design for systems evolving on a manifold. An observer can be designed for the two subsystems individually. 
	For the subsystem evolving on the Lie group, we 
	define a group action on the measurement as well. A constructive procedure for observer design is laid down for 
	this subsystem. In particular, we detail a gradient based observer design technique
	 stemming from a choice of suitable cost function, that makes the error dynamics 
	 autonomous. 
	However, for the system evolving on the base space, the methodology is not
	uniform and is implemented on a case by case basis. We do not examine this here.
		
	The initial ideas behind this theory of symmetry-preserving observers appear in \cite{bonnabel-2008}. We work in a similar setting as \cite{bonnabel-2008}. The same problem, if the configuration manifold $P$ itself is a Lie group, is presented in \cite{bonnabel-2009}. 
They, however, adopt a very different methodology from us. They use a method inspired from observer design for linear systems, like the idea of the Luenberger observer, in which they augment the vector field governing the original system with a correction (gain and innovation) term to have desirable characteristics of the error dynamics, and rely on linearization of the system to design the observer gains to obtain desirable characteristics.  In \cite{bonnabel-2008},  the natural decomposition of the system due to symmetry is stated briefly,  without proofs or much geometric insight. The work of Mahony {\it et. al.} \cite{mahony-2013,mahony-2018} is in the same setting as ours and \cite{mahony-2017}	applies the methodology in \cite{mahony-2013} to design an observer for the SLAM problem when the group action is {\it transitive.} If in our setting the group action is assumed to be transitive, we observe that the cross-section reduces to a single point on the configuration manifold, and there is exactly one orbit.  Hence given a base point on the manifold, the rest of the manifold can be identified with the Lie group.{(Consider the action of $G = \SOthree$ on 
	$P = \mathbb{S}^2$. The action is transitive i.e. given any element $p \in P$ we can obtain any other element $p' \in P$ via a suitable rotation i.e. the action of a suitable element in $\SOthree$. Mathematically, given any $p$ and $p'$ in $\mathbb{S}^2$, there exists $g \in \SOthree$ (non-unique) such that $Rp = p'$, see for example, Given's rotations \cite{watkins}. Hence, just one element of $P$ suffices to describe the entire space $P$ based on the action of $\SOthree$.)}
	It therefore essentially reduces to observer design on a Lie group, a particular case of the methodology we propose. 
	 {\it However,  all group actions may not always be transitive. If the group action is not transitive, there is the formation of quotient manifold and orbit space which we highlight in this current work. }	

	The two examples we present highlight many aspects of the underlying machinery. In the first one,  the action is
	 of $\SOthree$ on $\R^3 \setminus \{0\}$. This action is not free and has a non-trivial isotropy subgroup. We display the decomposition of $\R^3 \setminus \{0\}$ into the base and fibre co-ordinates, and the decomposition of its tangent space into the horizontal and vertical component. These demonstrations are instructive since the non-trivial isotropy subgroup makes the problem involved. We proceed to analyse kinematics of an object whose trajectory evolves on  $\R^3 \setminus \{0\}$. We conclude by showing how the well known problem of tracking the object using range and bearing measurements falls into our geometric structure of base and fibre coordinates. 
	The second example concerns the SLAM problem. Briefly, the problem here involves a vehicle in an unknown environment and the goal is to create a map of the environment and concurrently determine the location of the vehicle relative to the environment. SLAM has received significant attention in literature \cite{bailey-2006,cadena-2016,
	dissanayake-2011,durrant-whyte-2006,mahony-2017}. 
We begin with the geometric structure introduced in \cite{mahony-2017}, and proceed to study it in greater depth. 
The action of the Lie group on the configuration space in the SLAM problem is a free action, which leads to the formation of the the principal fibre bundle structure. We present the decomposition of the manifold into base and fibre coordinates and show that this problem admits a global cross-section. Studying the decomposition of the tangent space into vertical and horizontal spaces leads us to the decomposition of the SLAM kinematics system into two smaller subsystems. This examination of the geometric structure has not been elaborated in literature before. 
The sub-system evolving on the base space is observable, but the one evolving on the Lie group is not observable (this observation is made in  \cite{kwlee-2018,mahony-2017,wang-2018}). We end by showing that if we introduce a set of non-collinear but known landmarks, we can design an observer for the group using a known technique for observer design on $\SEthree$ \cite{mahony-2011}.

The paper is organised as follows. Section \ref{sec:mathprelim} takes a look at the fibre bundle structre, and highlights the decomposition of the manifold into base and fibre coordinates, created by the Lie group. These concepts are illustrated by showing an example of the action of $\SOthree$ on $\R^3 \setminus \{0\}$. 
Section \ref{sec:eqcs} addresses how equivariant control systems decompose in the presence of the preceding geometric structure into two smaller subsystems.
Section \ref{sec:liegroupobs} details a method to design an observer on a Lie group. Section \ref{sec:examples} concludes by presenting examples highlighting some of the developments in previous sections.

\section{Fibre Bundle Structure}
	\label{sec:mathprelim}
In this section we briefly introduce the reader to the mathematical tools that we employ
in the rest of the paper. 
Let $P$ be a smooth manifold of dimension $n_P$, $G$ be a $n_G$ dimensional connected Lie group, with $\id$ being the identity element, and let $\mathcal{G}$ be its Lie algebra. In what follows we will summarize several well known results that are crucial to this work such as group actions, orbit spaces, infinitesimal generators and invariance under the group action.

\subsection{Group actions and orbit spaces}
For any $h \in G$, let $L_h : G \to G$ denote the left multiplication. Let $\phi : G\times P\to P$ be a proper, constant rank, left action \footnote{for a right action, all results can be derived in an analogous fashion} of $G$ on $P$ and let
$\mathcal{X}_P$ denote the set of smooth vector fields on $P$. We will 
frequently use the notation $ g\cdot p  \triangleq \phi(g,p)$.
We distinguish between two maps - $\phi^p(\cdot): G \rightarrow P$ and
$\phi_g (\cdot): P \rightarrow P $ that are associated with $\phi(\cdot,\cdot)$ as follows:
\begin{align*}
	\phi_g(p) \triangleq \phi(g, p) \quad \forall g \in G \quad \text{and} \quad \phi^p(g) \triangleq \phi(g, p) \quad \forall p \in P.
\end{align*}

The orbit of $\phi$ through $p$ is defined to be the set of points
\begin{align*}
\mathcal{O}(p)&\triangleq \set{\phi_g(p)}{g\in G}.
\end{align*}
%
%

\begin{definition}
We take the following terminology from \cite[Chapter 9]{marsden}.
\begin{enumerate}
\item The group action is {\it free} if it has no fixed points, that is $\phi_g(p) = p$ implies $g = \id$ or equivalently, if for each $p \in P$, $g \mapsto \phi_g(p)$ is one-to-one. 
\item The group action is {\it transitive} if for all $p_1,p_2 \in P$ there exists $g \in G$ such that $p_2 = \phi_g(p_1)$. The manifold is therefore one single orbit of the group.
\item The group action is {\it proper} if the mapping
\begin{align*}
G \times P \ni (g,p)\mapsto (p,\phi(g,p)) \in P \times P
\end{align*}
is proper. 
\end{enumerate}
\end{definition}

\begin{figure}
\centering
\includegraphics[scale=0.8]{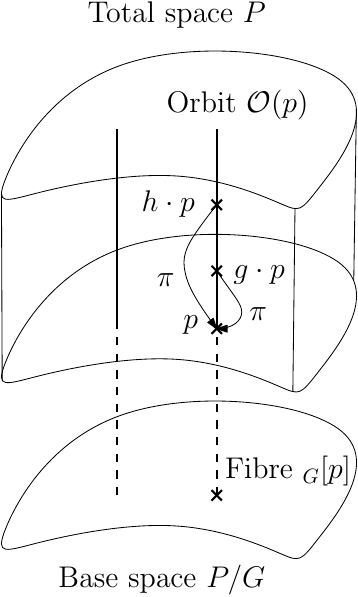}
\caption{Fiber bundle, projection, base space and orbits}
\label{fig:base-fibre}
\end{figure}

Since the orbits are equivalence classes we will also denote $\O(p)$ as $_G[p]$ 
\footnote{Given a smooth manifold $P$ and a Lie group $G$, we will denote the orbit of $p$ under $G$ (with left action) as $_G[p]$ and under right action as $[p]_G$. In particular, the smooth manifold $P$ may itself be a Lie group, and $G$ may be a Lie subgroup of $P$.}.  
Let $P/G$ denote the space of all orbits of $\phi$ with $\pi_\phi :P\to P/G$ denoting the canonical projection map. That is,  $\pi_\phi(p)=_G[p]$. {We will distinguish between $\O(p)$ and $_G[p]$ as follows : we will view $\O(p)$ as a collection of points in $P$ i.e. a subset of $P$ and $_G[p]$ as an element of $P/G$.} These elements can be visualised in Figure \ref{fig:base-fibre} which has been inspired from \cite[Figure 10.5.1]{marsden}.

Given $\zeta \in \mathcal{G}$, \emph{the infinitesimal generator} of the action is  the vector field $\zeta_P\in \mathcal{X}_P$ given by
\begin{align*}
\zeta_P(p)\triangleq \left.\dfrac{d}{dt}\right\rvert_{t=0}\phi_{\exp{\zeta t}}(p) =T_e\phi^p\cdot \zeta\quad \forall \quad p\in P
\end{align*}
Note that the flow of  $\zeta_P$  is  $\Phi_{\zeta_P}^t (p) \triangleq \phi_{\exp{\zeta t}}(p)$, where $\exp{\zeta t }\in G \quad \forall t \in \mathbb{R}$.
Since $T_e\phi^p$ is linear it follows that 
\begin{align*}
(\zeta+\eta)_P&=\zeta_P+\eta_P,\\
(\alpha\,\zeta)_P&=\alpha\,\zeta_P.
\end{align*}
%
By definition we also have
\begin{align*}	
		(\mathrm{Ad}_g \zeta)_{P}(p) = T_{e}  \phi^{p} 
		\cdot \mathrm{Ad}_g \zeta.
\end{align*}
Thus 
\begin{align*}	
		(\mathrm{Ad}_g \zeta)_{P}(\phi_g(p)) &= T_{e}  \phi^{\phi_g(p)} 
		\cdot \mathrm{Ad}_g \zeta\\
		&= \left.\dfrac{d}{ds}\right|_{s=0}\phi(g\exp{(\zeta s)}g^{-1},\phi(g,p))=\left.\dfrac{d}{ds}\right|_{s=0}\phi(g\exp{(\zeta s)},p)\\
		&=\left.\dfrac{d}{ds}\right|_{s=0}\phi_g\circ \phi_{\exp{(\zeta s)}} (p)=T_p\phi_g\cdot T_e\phi^p\cdot \zeta.
\end{align*}
This shows that 
\begin{align}
{T_p\phi_g\cdot \zeta_P(p)=T_p\phi_g\cdot (T_e\phi^p\cdot \zeta) =(\mathrm{Ad}_g\zeta)_P(\phi_g(p))\neq \zeta_P(\phi_g(p)),}\label{eq:AdEquivLeft}
\end{align}
and hence that in general $\zeta_P$ is not a $\phi$ - invariant vector field.
Differentiating this expression it also follows that 
\begin{align*}
-[\eta_P,\zeta_P]=[\eta,\zeta]_P,
\end{align*}
and thus for left actions the assignment $\zeta \to \zeta_P$ is a Lie algebra antimorphism and the subspace of vectorfields $\mathcal{X}_G\triangleq \{\zeta_P\,:\, \zeta\in \mathcal{G}\}$ is a Lie-subalgebra of the space of vectorfields $\mathcal{X}$ on $P$. Since this distribution is involutive, it is integrable. Since it is tangent to the orbits at every point of the orbit these integral manifolds are in fact the orbits of the action. The space $T_p\mathcal{O}(p)=\mathcal{X}_G(p)$ is called the \emph{vertical space} of the tangent space $T_pP$ of $P$ at $p$.

\begin{figure}
\centering
\includegraphics[scale=0.5]{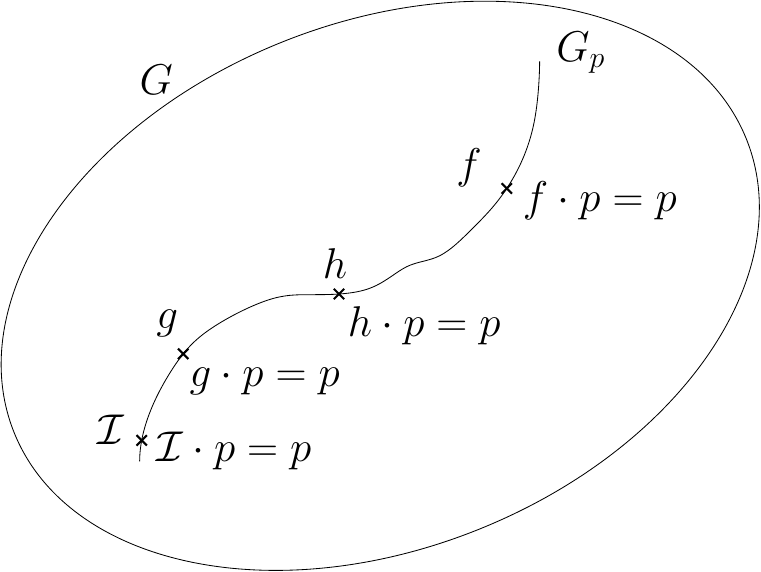}
\caption{Isotropy subgroup of $p$, $G_p \subset G$}
\label{fig:isotropy}
\end{figure}

Let $G_p\subset G$ be the isotropy subgroup of $p$. That is let $G_p\triangleq \set{g \in G}{ g\cdot p=p }$. The isotropy subgroup can be visualised in Figure \ref{fig:isotropy}. Then we see that $G_{g\cdot p}=gG_pg^{-1}$ and that $\zeta_P(p)=0$ for all $\zeta \in \mathcal{G}_p\triangleq T_eG_p$. Hence we have that the  $\mathcal{X}_G$ is rank $(\mathrm{dim}(G)-\mathrm{dim}(G_p))$ at $p$. The integral submanifolds of $\mathcal{X}_G$ coincide with the orbits of the group action. When the action is proper these integral manifolds are guaranteed to be embedded submanifolds. The properness also implies that
$P/G$ is Hausdorff and hence is a smooth manifold of dimension $\mathrm{dim}(P)-(\mathrm{dim}(G)-\mathrm{dim}(G_p))$ with respect to the usual quotient topology. This intutive picture is summarised in the following result.

\begin{theorem}[{\cite[Theorem 2.3.3]{kolk}}]
If the action $\phi : P\times G\to P$ is proper and constant rank then each $\pi^{-1}_\phi(_G[p])$ is a closed embedded submanifold of $P$ of dimension $(\mathrm{dim}(G)-\mathrm{dim}(G_p))$. Furthermore $P/G$ is a smooth manifold of dimension $r=\left(\mathrm{dim}(P)-(\mathrm{dim}(G)-\mathrm{dim}(G_p))\right)$. 
\end{theorem}

Now we move towards decomposing $P$ into the base and fibre coordinate. To that end, we present the definitions of the maps that will yield the coordinates, and then provide insight into how we obtain the coordinates.
\begin{definition}
Define $\sigma_P : P/G \to P$ to be a map that assigns to every element $_G[p] \in P/G$,
a point on the fiber $\mathcal{O}(p)$ in a smooth fashion. That is
	\begin{align*}	
	\sigma_P \;\;\; \text{is smooth and } \;\;\;\; \pi_\phi \circ\sigma_P =id_{P/G}.	
	\end{align*}
	Such a $\sigma_P$ is called a {\it { global} smooth cross section of $(P,P/G,\pi_\phi)$.}
\end{definition}
{
\begin{assumption}
We assume that such a $\sigma_P$ exists.
\end{assumption}
}

\begin{definition}
\label{def:gamma}
Based on the above fact, we define $\gamma_{\sigma_P} : P  \to \bigcup\limits_{z \in \sigma_P(P/G)} G/G_{z}$ such that $P \ni p \mapsto [g]_{{G_{\sigma_P(_G[p])}}} \in G/G_{\sigma_P(_G[p])}$ such that (\ref{eq:FibreCoordinates}) holds for all $g \in [g]_{{G_{\sigma_P(_G[p])}}}$. In other words, $
\gamma_{\sigma_P} (p) := \set{ g \in G}{p = g\cdot \sigma_P(_G[p]) } $.
\end{definition}

\begin{figure}
\begin{subfigure}{0.48\textwidth}
\includegraphics[scale=0.4]{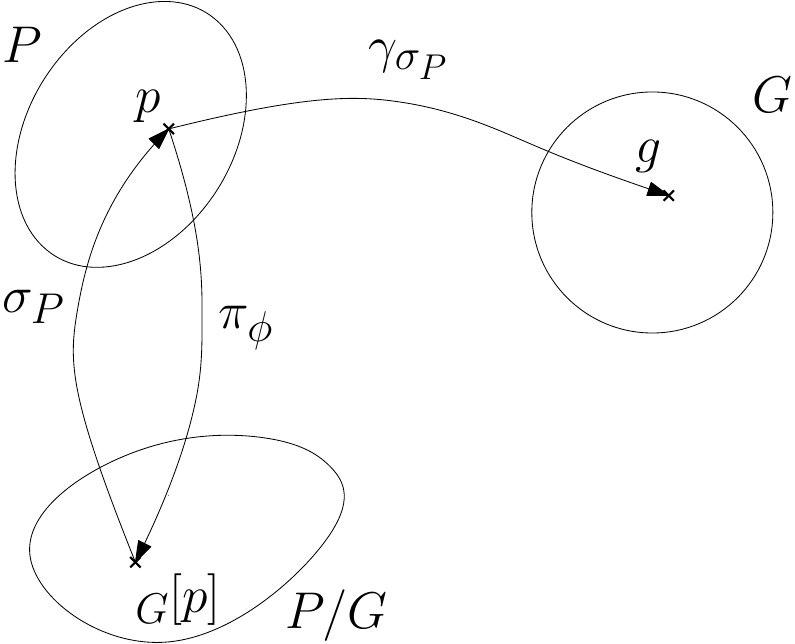}
\caption{Section assignment for the free group action}
\label{fig:section-free}
\end{subfigure}
\begin{subfigure}{0.48\textwidth}
\includegraphics[scale=0.4]{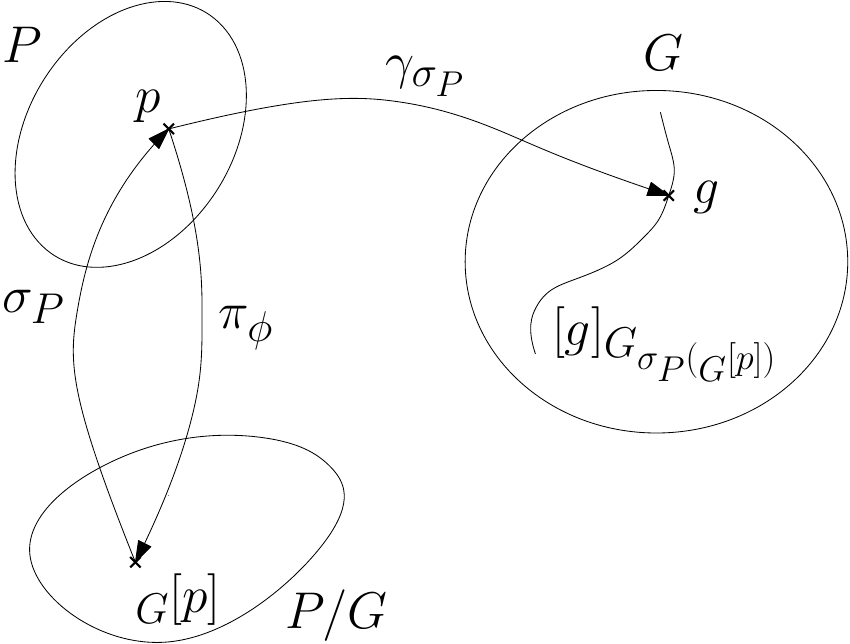}
\caption{Section assignment for the general case}
\label{fig:section-non-free}
\end{subfigure}
\caption{Section}
\end{figure}

Note that associated with the section $\sigma_P(\cdot)$ there exists  for every $p \in P$ a  $g\in G$ such that 
\begin{align}
p=\phi_g(\sigma_P(_G[p]))=\phi^{\sigma_P(_G[p])}(g)\eqqcolon g\cdot \sigma_P(_G[p])
\label{eq:FibreCoordinates}
\end{align}
Since $p=\phi_g(\sigma_P(_G[p]))=\phi_{gh}(\sigma_P(_G[p]))$ for all $h\in G_{\sigma_P(_G[p])}$ we see that  
the $g\in G$ that satisfies the above relationship is unique only up to a right multiplication by an element of $G_{\sigma_P(_G[p])}$. 
That is if $g_{p}\in G$ is such that 
$p=\phi_{g_p}(\sigma_P(_G[p]))$ and $[g_p]_{G_{\sigma_P(_G[p])}}\triangleq \set{g_ph}{ h\in G_{\sigma_P(_G[p])}}$ then $[g_p]_{G_{\sigma_P([p])}}\in G/G_{\sigma_P([p])}$ can be uniquely identified with $p\in \mathcal{O}(p)$.
Thus we see that there exists a unique $[g_p]_{G_{\sigma_P(_G[p])}}\in G/G_{\sigma_P(_G[p])}$ such that (\ref{eq:FibreCoordinates}) holds for all $g\in [g_p]_{G_{\sigma_P(_G[p])}}$ and hence that the cross section $\sigma_P$ allows us to identify points in $_G[p]=\mathcal{O}(p)$ with points in $G/G_{\sigma_P([p])}$ in a unique way.
Observe that $\sigma_P(_G[\phi_h(p)])=\sigma_P(_G[p])$ for all $h\in G$. Thus using the expression (\ref{eq:FibreCoordinates}) we see that 
\begin{align}
\gamma_{\sigma_P}(\phi_h(p))=\bar{L}_h\gamma_{\sigma_P}(p)
\label{eq:Lbar}
\end{align}
where $G/G_z \ni [g]_{G_z} \mapsto \bar{L}_h([g]_{G_z}) := \set{h\cdot g'}{g' \in [g]_{G_z}} = [hg]_{G_z} $ for any $z \in P$.
That is the following commutative diagram holds.
\[ 
\begin{tikzcd}
P \arrow{r}{\phi_h} \arrow[swap]{d}{\gamma_{\sigma_P}} & P  \arrow{d}{\gamma_{\sigma_P}}   \\%
G/G_{\sigma_P(_G[p])}  \arrow{r}{\bar{L}_h}& G/G_{\sigma_P(_G[p])}
\end{tikzcd}
\]
Also observe that $\gamma_{\sigma_P}(\sigma_P(_G[p])) = [\id]_{G_{\sigma_P(_G[p])}}$. 
\begin{remark}
{ In summary, given any $p \in P$, we identify it using $z \coloneqq \sigma_P(p)$ and $[g_p]_{G_z} \coloneqq \gamma_{\sigma_P}(p)$ such that $p = g \cdot z$ for all $g \in [g_p]_{G_z}$.} The concept of the section can be visualised in Figure \ref{fig:section-free} for the free action and Figure \ref{fig:section-non-free} for the general case.
\end{remark}
\begin{remark}
	Notice that the map $\gamma_{\sigma_P}$ depends on the cross section $\sigma_P : P/G \to P$. In the special case where the action is transitive picking the cross section simply amounts to identifying a particular point $p_\sigma\in P$ and then we see that $P\simeq G/G_{p_\sigma}$.
\end{remark}

\begin{definition}
The coordinate $\sigma_P(p)$, or equivalently $_G[p]$, is called the base coordinate of $p$ while $[g]_{G_{\sigma_P(_G[p])}}$ is called the fibre coordinate of $p$.
\end{definition}

\begin{remark}
If the action $\phi : P\times G\to G$ is free then the bundle $(P,G,\pi_\phi)$ is a principal fibre bundle \cite{kobayashi}.
\end{remark}
Let us give some more attention to the case of the free action and the transitive action. 

If the action of $G$ on $P$ is free,
 then for every $p \in P$, there exists a unique $g \in G$ such that \eqref{eq:FibreCoordinates} holds, that is, 
\begin{align}
p = g \cdot \sigma_P(_G[p]) 
\label{eq:FibreCoordinates-free} 
\end{align}
This leads to the definition of $\gamma_{\sigma_P} : P \to G$ as $\gamma_{\sigma_P}(p) \coloneqq g$ where $g$ satisfies \eqref{eq:FibreCoordinates-free}. The relation \eqref{eq:Lbar} is simplified to $\gamma_{\sigma_P}(\phi_h(p)) = L_h(\gamma_{\sigma_P}(p))$. 

If the action of $G$ on $P$ is transitive, the entire manifold is a single orbit of the group. Therefore, we can pick arbitrary $p_{\sigma} \in P$ such that for any $p \in P$, there exists a $g \in G$ such that 
\begin{align}
p = \phi_g(p_{\sigma})
\label{eq:FibreCoordinates-tran} 
\end{align}
This immediately yields a choice of section as $\sigma_P(p) = p_{\sigma}$ for all $p \in P$, since the base space consists of a single point. Observe that given any $h \in G_{p_\sigma}$, $gh$ also satisfies \eqref{eq:FibreCoordinates-tran}. This leads to the definition of $\gamma_{\sigma_P} : P \to G/G_{p_{\sigma}}$ as $\gamma_{\sigma_P}(p) \coloneqq [g]_{G_{p_{\sigma}}}$ where $g$ satisfies \eqref{eq:FibreCoordinates-tran}. 

\subsection{Decomposition of Tangent Space}

	The group action leads to the decomposition of the tangent space. Let us look at the case of free group action first, since it will natually lead us to the case when the group action is not free. 	
	At any point $p \in P$, $T_pP$ decomposes into two complementary vector spaces - the vertical space, denoted $\Ver_p(P)$ and horizontal space, denoted $\Hor_p(P)$. The vertical space($\Ver_p(P)$) at each point is tangent space to the orbit passing through that point, and is therefore is isomorphic to the Lie algebra of $G$. It is also the kernel of $T_p\pi_{\phi}$. The horizontal space ($\Hor_p(P)$), is non-unique and can be chosen by the user to satisfy $\Hor_p(P)\bigoplus\Ver_p(P) = T_pP$ and $\Hor_{g \cdot p}(P) = T_p\Phig{g}\Hor_p(P)$. It can be easily shown that every point, $T_p \gamma_{\sigma_P}$ annihilates horizontal vectors and $T_p\pi_{\phi}$ annihilates vertical vectors, and $T_{_G[p]}\sigma_P$ is a bijection between  $T_{_G[p]}P/G$ and $\Hor_{\sigma_P(p)}(P)$. For more details, see \cite{kobayashi}.

Proceeding to the case when the group action is not free, the vertical and horizontal space still have the same definition and properties. However, it is instructive to observe that the vertical space has two distinct subspaces. Let a basis for the Lie algebra of $G$ be $\mathcal{B}_1 = \{b_1,b_2,\ldots,b_n\}$, and let the basis for the Lie algebra of $G_p$ be $\mathcal{B}_2 = \{b_1,b_2,\ldots,b_m\} \subset \mathcal{B}_1$ with $m < n$. Then $\mathcal{B}_2$ represents all those directions, going along which, one remains in the stabiliser subgroup $G_p$, therefore, the curves that they induce on $P$ via the tangent map of $\phi^p$ will have zero tangent vector. 
	
\begin{figure}
\centering
\includegraphics[scale=0.8]{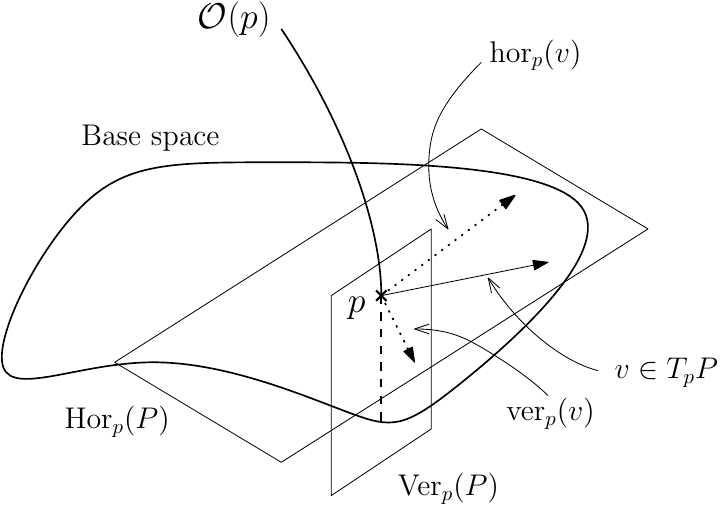}
\caption{Horizontal and vertical space decomposition at any arbitrary point $p \in P$.}
\label{fig:hor-ver}
\end{figure}

	Given a tangent vector $v_p \in T_p P, \ver(v_p)$ will denote its vertical component and $\hor(v_p)$ will denote its horizontal component. An illustration of this is in Figure \ref{fig:hor-ver}.
	
We illustrate some of these ideas using a well known example.
\begin{example}
\label{eq:SOonR3}

Consider the action of the rotation group $\SOthree$ on the elements of $ \R^3\setminus \{0\}$ given by
\begin{align*}
\phi : \SOthree\times (\R^3\setminus \{0\}) \to (\R^3\setminus \{0\} ) \quad (R, p) \to Rp
\end{align*}
Define the orbit of $p$ as:
\begin{align*}
\O(p) = \set{Rp}{R \in \SOthree} 
\end{align*}
The collection of all such orbits in $\mathbb{R}^3\setminus \{0\}$ is the quotient space: %
\begin{align*}
(\mathbb{R}^3 \setminus\{0\}) /\SOthree = \set{_{\SOthree}[p]}{p \in \R^3\setminus \{0\} }
\end{align*}
We define a projection:
\begin{align*}
\pi : \R^3\setminus \{0\} \to (\R^3\setminus \{0\} / \SOthree ) \quad  \pi(p) = _{\SOthree}[p]
\end{align*}
Let $||\cdot||_2$ denote the Euclidean norm on $\R^3$. Since all the orbits are spheres of radius $||p||_2$ we can identify all points in $(\mathbb{R}^3 \setminus\{0\})/\SOthree$ with points in $\R_{> 0}$, and it also helps in defining the section as follows.Choose an arbitrary point in $\R^3\setminus \{0\} $. Let that point be $p_0 = (1, 0, 0)$. 
\begin{align*}
{\sigma_P} : (\mathbb{R}^3 \setminus\{0\})/\SOthree \to (\mathbb{R}^3 \setminus\{0\}) \quad {\sigma_P}([p]) :=||p||_2(1, 0 , 0)
\end{align*}

This corresponds to a smooth section of the bundle. $||p||_2(1,0,0)$ is then termed the base co-ordinate of $p$. 

{\bf Fact}: Given  any $p  \in \R^3\setminus \{0\}$ there exist Given's rotations $R_1,R_2 \in \SOthree$ such that $R_2 p = ( *, * , 0 )^T$ and $R_1 R_2 p = ( ||p||_2 , 0 , 0 )^T$ \cite[Chapter 3]{watkins}. 

To get the fibre co-ordinate, define 
\begin{align*}
\gamma_{\sigma_P} : (\mathbb{R}^3 \setminus\{0\})  \to \bigcup\limits_{z \in \sigma_P(\mathbb{R}^3 \setminus\{0\})/\SOthree)} \SOthree/G_{z}
\end{align*}
$\sigma_P(\mathbb{R}^3 \setminus\{0\}) = \set{\lambda p_0}{\lambda \in \R_{>0}}$, and for any $z \in \sigma_P(\mathbb{R}^3 \setminus\{0\})$, $G_z = G_{p_0}$. The definition simplifies to $\gamma_{\sigma_P} : (\mathbb{R}^3 \setminus\{0\}) \to \SOthree/G_{p_0}$ defined as $\gamma_{\sigma_P}(p) \coloneqq [R_2^T R_1^T]_{G_{p_0}}$. 

Thus $p=(||p||_2, R_2^T R_1^T)$ and $||p||_2$ is called the base coordinate and $[R_2^T R_1^T]_{G_{p_0}}$ is called the fibre coordinate of $p$.
To characterise the vertical space at $p \in \R^{3} \setminus \{0\}$, we will find the null space of $T_p \pi$.
Consider a smooth curve $r(t)R(t) e_1  \in \R^3\setminus \{0\}$ where $R(\cdot)$ is a smooth curve in $\SOthree$ with $R(0)= R_{1} R_{2}$,
$\dot{R}(0) = [R(0)]\Omega_{\times} \in \mathfrak{so(3)}$, and $r(t)$ is a smooth curve in $\R^3$ , $r(0) = ||p||$ and $\dot{r}(0) \in \R$. Then 
\begin{align*}
	r(0) R(0) e_1 = ||p|| R_{1} R_{2} e_1 = p 
\end{align*}
\begin{align*}
  v = \dot{r}(0) R_{1}R_{2} e_1 + 
				||p||\dot{R}(0) e_1 =  \dot{r}(0) R_{1}R_{2} e_1 + 
				||p|| R \Omega_{\times}  e_1 
\end{align*}

Let $\dot{r}(t) = 0$ so that $v \in T_p \pi$.
Then it is evident that $p^Tv = 0$. Thus, $\Ver_p(\R^{3} \setminus \{0\}) = \set{v \in \R^3}{p^Tv = 0}$. And we choose the horizontal space as $\Hor_p(\R^{3} \setminus \{0\}) = \set{\lambda p}{\lambda \in \R}$. Therefore, given $v \in T_p(\R^{3} \setminus \{0\})$, $\ver(v) = v - (p^Tv)\frac{p}{r^2}$ and $\hor(v) = (p^Tv)\frac{p}{r^2}$.
\end{example}

\section{Equivariant Control Systems}
\label{sec:eqcs}

Let $P$ (state-space), $\mathcal{U}$ (control input), and $\mathcal{Y}_1$, $\mathcal{Y}_2$ (measurements) be $n_P,n_{\mathcal{U}},n_{\mathcal{Y}_1},n_{\mathcal{Y}_2}$ - dimensional smooth manifolds respectively and let $X: {P}\times \mathcal{U} \to T{P}$ be a smooth map such that $X(p,u)\in T_p{P}$ for all $p\in{P}$ and $u\in\mathcal{U}$ and $H: P\to \mathcal{Y}_2$ is smooth and onto. 
Let $\phi : G\times P \to P$, $\psi : G\times \mathcal{U} \to \mathcal{U}$ be proper and constant rank left actions of $G$ on $P$ and $\mathcal{U}$ respectively, and let ${\varphi} : G\times \mathcal{Y}_1 \to \mathcal{Y}_1$ and $\oldtilde{\varphi} : G\times \mathcal{Y}_2 \to \mathcal{Y}_2$  be proper left actions of $G$. Define the proper constant rank left action $\rho :G\times (P\times \mathcal{U}) \to P\times \mathcal{U}$ by $\rho_g(p,u)=(\phi_g(p),\psi_g(u))=(\phi_g\times \psi_g)(p,u)$. Define the section associated to $\pi_{\phi}$ as $\sigma_P$ and additionally define $\gamma_{\sigma_P}$ associated with $\sigma_P$. Define $P \supset \K \coloneqq \sigma_P(P/G)$. Let $y_0 \in \Y_1$ be fixed and known.
Based on these structures, the equations 
\begin{align}
\dot{p}&=X(p,u),\label{eq:DynamicSystem}\\
y_{G} &=\varphi_{g^{-1}}(y_0) \\
y_{\K}&=H(p),\label{eq:Output} 
\end{align}
define a control system with state evolving on $P$ with control taking values in $\mathcal{U}$ and the output taking values in $\mathcal{Y}$.
The 6-tuple $(P,\mathcal{U},\mathcal{Y}_1,\mathcal{Y}_2, X, H)$ will be referred to as a \emph{Control System} on $P$.
Since the notion of symmetries plays an important role in our evolution of ideas, we define group actions.
\begin{definition}
A smooth onto map $H:P\to \mathcal{Y}_2$ is said to be $G$ - equivariant if
$\oldtilde{\varphi}_g\left(H(p)\right)=H\left(\phi_g(p)\right)$ for all $p\in P$ and $g\in G$.
That is if the following commutative diagram holds.
\[ 
\begin{tikzcd}
P \arrow{r}{H}\arrow[swap]{d}{\phi_g}  & \mathcal{Y}_2 \arrow{d}{\oldtilde{\varphi}_g}  \\
P \arrow{r}{H}  & \mathcal{Y}_2  
\end{tikzcd}
\]
\end{definition}

\begin{definition}
The control system $(P,\mathcal{U},\mathcal{Y}_1,\Y_2,X,H)$ will be called a $G$ - equivariant control system if the maps $X : {P}\times \mathcal{U} \to T{P}$ and $H:{P}\to \mathcal{Y}_2$ are G - equivariant. That is, if the following two commutative diagram holds.
\[ 
\begin{array}{cc}{\quad\begin{tikzcd}
P\times\mathcal{U} \arrow{r}{X}  \arrow[swap]{d}{\phi_g\times \psi_g}& TP  \arrow{d}{T\phi_g} \\
P\times\mathcal{U} \arrow{r}{X} &TP
\end{tikzcd}} \quad &\quad 
{\begin{tikzcd}
P \arrow{r}{H}  \arrow[swap]{d}{\phi_g}& \mathcal{Y}_2 \arrow{d}{\oldtilde{\varphi}_g} \\
P \arrow{r}{H} &\mathcal{Y}_2
\end{tikzcd}}
\end{array}
\]
\label{invconsys}
\end{definition}

We now define projection maps on each of the spaces to impart a bundle structure to each. 
Let $\pi_{\phi} :P \to P/G$,  $\pi_{\psi} :\mathcal{U} \to \mathcal{U}/G$,  
and $\pi_{\rho} :P\times \mathcal{U} \to (P\times \mathcal{U})/G$ be the respective canonical projections.
Denote by $_G[p]  (\in P/G) =\pi_\phi(p)$ the orbit of the $\phi$ action of $G$ through $p$, $_G[u]=\pi_\psi(u)$ the orbit of the $\psi$ action of $G$ through $u$ ($\in \mathcal{U}/G$),
and $_G[p,u]=\pi_\rho(p,u)$ the orbit of the $\rho$ action of $G$ through $(p,u)$. Note that, in general, $_G[p,u]\neq (_G[p],_G[u])$ unless $\psi_g$ is the identity map, in which case $_G[p,u]= (_G[p],u)$.

We will demonstrate below the known result that the flow of $G$ - equivariant control systems take orbits to orbits. 
Let $p(t)\triangleq \Psi^X_{t} (p_0,u([0,t]))$ be the solution of (\ref{eq:DynamicSystem}) for a control history $u([0,t])$.
Consider the curve
$\phi_g(p(t))$ for some $g\in G$. Then from $G$ - invariance we have
\begin{align*}
\dfrac{d}{dt}\phi_g\left(p(t)\right)&=T_{p(t)}\phi_g\cdot X\left(p(t),u(t)\right)= X\left(\rho_g(p(t),u(t))\right).
\end{align*}
Thus we have that $\phi_g(p(t))$ is the solution of $X$ that originates at $\phi_g(p_0)$ with control history $\psi_g(u(\cdot))$. Thus it follows that  for any $p_0\in P$
\begin{align*}
\Psi^X_{t}\left(\rho_g(p_0,u([0,t]))\right)=\phi_g\left(\Psi^X_{t}(p_0,u([0,t]))\right)\quad \forall \quad t \geq 0,\:\text{and} \: g\in G. 
\end{align*}
and hence
\begin{align*}
\pi_\phi\left(\Psi^X_{t}(p_0,u([0,t]))\right)=\pi_\phi\left(\phi_g\left(\Psi^X_{t}(p_0,u([0,t]))\right)\right)=\pi_\phi\left(\Psi^X_{t}\left(\rho_g(p_0,u([0,t]))\right)\right).
\end{align*}
That is $\pi_\phi\circ \Psi^X_{t} (p',u'([0,t]))= \pi_\phi\circ \Psi^X_{t} (p_0,u([0,t]))$ for all $p'=\phi_g(p_0)$ and $u'(\cdot)=\psi_g(u(\cdot))$ for all $g\in G$ and hence that the flow of $G$ - equivariant control systems take orbits to orbits.

We are now in a position to describe the evolution on the base
space $P/G$, and present a differential equation on this space.
Let us define a smooth map $\bar{X} : ({P}\times \mathcal{U})/G \ra T(P/G)$ such that the following commutative diagram holds:
\[ 
\begin{tikzcd}
{P}\times \mathcal{U} \arrow{r}{\pi_{\rho}} \arrow[swap]{d}{X} & ({P}\times \mathcal{U})/G  \arrow{d}{\bar{X}}   \\%
TP  \arrow{r}{T\pi_{\phi}}& T(P/G)
\end{tikzcd}
\]
Therefore, $\bar{X} \circ \pi_{\rho} = T\pi_{\phi} \circ X$. 

\begin{claim}
$\bar{X}$ thus evaluated is a well defined map, and yields the same result irrespective of the particular point $(p,u) \in {P}\times \mathcal{U}$ on the orbit $_G[p,u]$ which is chosen at which to evaluate $X$ and $T_p\pi_{\phi}$.
\end{claim}
\begin{proof}
Assume that $(\phi_g(p),\psi_g(u))$ is another point on $_G[p,u]$ for some $g \in G$. Then if we evaluate $\bar{X}$ using this point we have $\bar{X}(_G[p,u]) = T_{\phi_g(p)}\pi_{\phi}\cdot X(\phi_g(p),\psi_g(u)) = T_{\phi_g(p)}\pi_{\phi}( T_p\phi_g\cdot X(p,u)) = T_p(\pi_{\phi} \circ \phi_g) \cdot X(p,u)$. Since $\pi_{\phi} \circ \phi_g = \pi_{\phi}$ we have that $\bar{X}(_G[p,u]) = T_p\pi_{\phi}\cdot X(p,u)$. Hence $\bar{X}$ is a well defined map.
\end{proof}

Hence given any $_G[p,u] \in ({P}\times \mathcal{U})/G $, $\bar{X}(_G[p,u]) = T_p\pi_{\phi}\cdot X(p,u)$.

Since $_G[p](t)=\pi_\phi(p(t))$, we have that
\begin{align*}
\dfrac{d}{dt}_G[p]&=T_p\pi_\phi \cdot X(p,u) = \bar{X}(_G[p,u]),
\end{align*}
does not depend on $g\in G$. 

Recall from Definition \ref{def:gamma} the map $\gamma_{\sigma_P} : P  \to \bigcup\limits_{z \in \sigma_P(P/G)} G/G_{z}$  takes $p\mapsto [g]_{{G_{\sigma_P(_G[p])}}}$ such that the relationship (\ref{eq:FibreCoordinates}) holds. \emph{Note that this map depends on the cross section $\sigma_P$.}
Also recall that $\gamma_{\sigma_P}\circ\phi_h=\bar{L}_h \circ \gamma_{\sigma_P}$ for all $h\in G$. %

\begin{figure}
\centering
\includegraphics[scale=0.8]{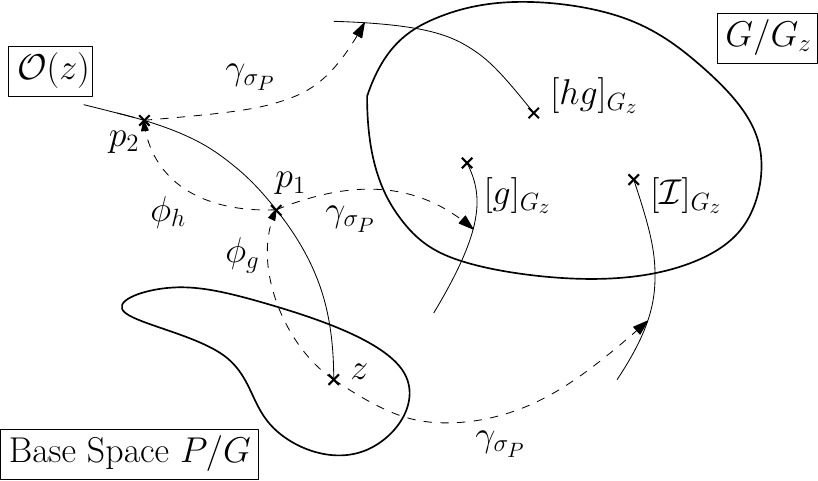}
\caption{Action of $\gamma_{\sigma_P}$}
\label{fig:gamma-sigma-p}
\end{figure}

\begin{remark}
Note that $\gamma_{\sigma_P}$ represents a family of coordinates on $P$, which when restricted to a single fibre, assigns respective points on the modulo space $G/G_z$ to every point on that fibre (where $z$ is the image of that fibre under $\sigma_P$). See Figure \ref{fig:gamma-sigma-p} for an illustration. 
\end{remark}

The next lemma presents
the differential equation for the evolution of the equivalence class  $\gamma_{\sigma_P}(p)$.

\begin{figure}
\centering
\includegraphics[scale=0.7]{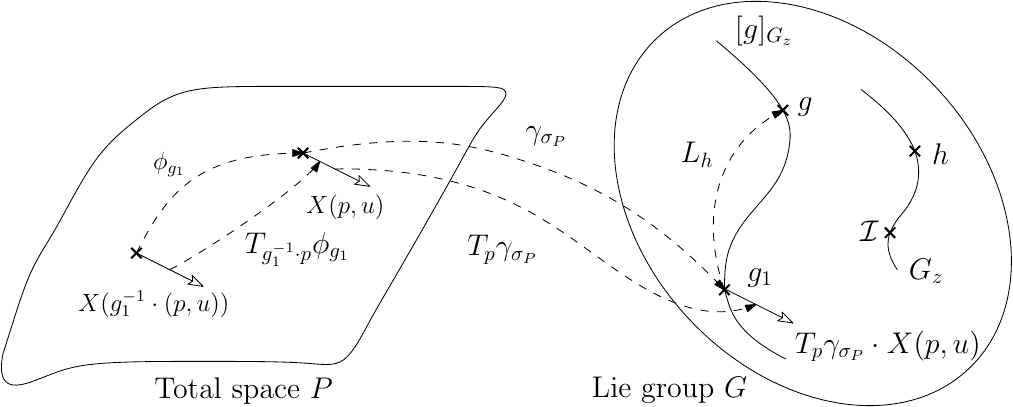}
\caption{Figure for the proof of Lemma \ref{lemma:group}. (Arrows indicate vectors)}
\label{fig:group-lemma}
\end{figure}

\begin{lemma}
Let $z \coloneqq \sigma_P(p)$.
The maps $\sigma_P(\cdot)$ and $\gamma_{\sigma_P}(\cdot)$ determine the dynamics of $\gamma_{\sigma_P}(p) = [g]_{G_z}$ as
\begin{align} \label{eq:dyn-fib}
\dfrac{d}{dt} {[g]_{G_z}}= T_{[\id]_{G_z}}\bar{L}_g \left( T_{z}\gamma_{\sigma_P} \left( X(z,\psi_{g^{-1}}(u)) - T_{_G[p]}\sigma_P(T_p\pi_{\phi} \cdot X(p,u)) \right) \right)
\end{align}
where $g \in [g]_{G_z}$. If $\psi_g$ is identity, then the dynamics are independent of the particular $g \in [g]_{G_z}$ that is chosen.
\label{lemma:group}
\end{lemma}
%
\begin{proof}
Let $g(\cdot)$ be such that $g(t) \cdot z(t) = p(t)$ for all time. Choose any element $h \in G_z$, then $h \cdot z = z$ which gives $h^{-1} \cdot z = z$, and $gh \in [g]_{G_z}$. Note that $g \in [g]_{G_z}$ also (if $h = \id$). Also note that $(\phi_h \circ \sigma \circ \pi_{\phi})  (p) = (\sigma \circ \pi ) (p)$, since $h$ is the stabiliser of $z$. Let $g_1(t) := g(t)h$ which implies $z(t) = (g_1(t))^{-1}p(t)$.
The rate of change of the fibre coordinate is given by the vertical component of the dynamics (after transferring it to the appropriate tangent space via appropriate tangent maps). Therefore,
\begin{align*}
\dfrac{d}{dt} {[g]_{G_z}}&=T_p\gamma_{\sigma_P}\cdot \ver_p(X(p,u)),\\
&= T_p\gamma_{\sigma_P} (X(p,u) - \hor_p(X(p,u))), \\
\end{align*}
Recall that $T_p\pi$ annihilates vertical vectors, $T_{_G[p]}\sigma_P$ is a bijection between  $T_{_G[p]}P/G$ and $\Hor_{\sigma_P(p)}(P)$ and $T_p\phi_g$ is used to transport horizontal vectors between tangent spaces. This gives us
\begin{align*}
\dfrac{d}{dt} {[g]_{G_z}}&= T_p\gamma_{\sigma_P} \left(X(p,u) - T_z\phi_{g_1}(T_{_G[p]}\sigma_P(T_p\pi_{\phi} \cdot X(p,u))) \right), \\
&= T_p\gamma_{\sigma_P}\left(T_z\phi_{g_1} \left( X(z,\psi_{g^{-1}}(u)) - T_{_G[p]}\sigma_P(T_p\pi_{\phi} \cdot X(p,u)) \right) \right), \\
&= T_{z} \left(\gamma_{\sigma_P} \circ \phi_{g_1}\right) \cdot\left( X(z,\psi_{g^{-1}}(u)) - T_{_G[p]}\sigma_P(T_p\pi_{\phi} \cdot X(p,u)) \right),\\
\end{align*}

Recalling \eqref{eq:Lbar} we see that 
\begin{align*}
\dfrac{d}{dt} {[g]_{G_z}}&=T_{z} \left(\bar{L}_{g_1} \circ  \gamma_{\sigma_P} \right) \cdot \left( X(z,\psi_{g^{-1}}(u)) - T_{_G[p]}\sigma_P(T_p\pi_{\phi} \cdot X(p,u)) \right),\\
&=T_{[\id]_{G_z}}\bar{L}_{g_1} \left( T_{z}\gamma_{\sigma_P} \left( X(z,\psi_{g^{-1}}(u)) - T_{_G[p]}\sigma_P(T_p\pi_{\phi} \cdot X(p,u)) \right) \right)
\end{align*}
See Figure \ref{fig:group-lemma} for an illustration of the steps below.
Choosing $h=I$ yields the first part of the proposition. Now assume that $\psi_g$ is identity. Then 
\begin{align*}
(T_{[\id]_{G_z}}\bar{L}_{g_1} \circ T_{z}\gamma_{\sigma_P}) \cdot X(z,\psi_{g^{-1}}(u))   &=
T_{z}(\gamma_{\sigma_P} \circ \phi_{g_1}) \cdot( X(z,\psi_{g^{-1}}(u)), \\
&= T_{z} \left(\gamma_{\sigma_P} \circ \phi_{g_1}\right) \cdot X\left(\phi_{g_1^{-1}}(p),u\right),\\
&=T_{z} \left(\gamma_{\sigma_P} \circ \phi_{g_1}\right) \cdot X\left(\phi_{h^{-1}} \circ \phi_{g^{-1}}(p),u\right),\\
&=T_{z} \left(\gamma_{\sigma_P} \circ \phi_{g_1}\right) \cdot X\left(\phi_{h^{-1}} (z),u\right),\\
&= T_{z} \left(\gamma_{\sigma_P} \circ \phi_{g_1}\right) \cdot (T_z\phi_{h^{-1}} \cdot X\left((z),u\right)),\\
&=T_{z} \left(\gamma_{\sigma_P} \circ \phi_{g}\right) \cdot X\left(z,u)\right),\\
&=T_{z} \left(\bar{L}_{g} \circ  \gamma_{\sigma_P} \right) \cdot X\left(z,u \right),\\
&=T_{[\id]_{G_z}}\bar{L}_{g} ( T_{z}\gamma_{\sigma_P} \cdot X\left(z,u \right))
\end{align*}
which is independent of the particular $h$ chosen, they are independent of the particular representative element in $[g]_{G_z}$ chosen. Using a similar method, and noting along the way that 
\begin{align*}
T_p\pi_{\phi} \cdot X(p,u) =  T_z(\pi_{\phi} \circ \phi_{g_1}) \cdot X(z,u) = T_z\pi_{\phi}  \cdot X(z,u)
\end{align*}
it can be shown that 
\begin{align*}
T_{[\id]_{G_z}}\bar{L}_g \left( T_{z}\gamma_{\sigma_P} \left(  T_{_G[p]}\sigma_P(T_p\pi_{\phi} \cdot X(p,u)) \right) \right)
\end{align*}
is independent of the particular $h$ chosen, they are independent of the particular representative element in $[g]_{G_z}$ chosen.

Therefore, the dynamics are independent of the particular $h$ chosen, they are independent of the particular representative element in $[g]_{G_z}$ chosen.
\end{proof}

\begin{remark}
In observer design, the control $u(\cdot)$ are {\it open loop} since they are given to the user and known. Therefore, the kinematics $X$ are essentially a function of time in the second argument. $\psi_g$ being identity is a valid function in this case. A similar observation has also been noted in \cite{bonnabel-2008} just before Definition 2. Often, $u(t)$ may represent velocity or actuator measurements.
\end{remark}

We present corollaries to Lemma \ref{lemma:group} when the action of $G$ on $P$ is free or transitive. They are an easy consequence of Lemma \ref{lemma:group} when we recall that $T_{_G[p]}\sigma_P(T_p\pi_{\phi} \cdot X(p,u))$ is a horizontal vector, and $T_{z}\gamma_{\sigma_P}$ annihilates horizontal vectors in these cases.
\begin{corollary}
If the action of $G$ on $P$ is free, defining $z \coloneqq \sigma_P(p)$ the dynamics of $g \coloneqq \gamma_{\sigma_P}(p)$ evolve as
\begin{align*}
\dfrac{d}{dt} {g}= T_{\id}{L}_g ( T_{z}\gamma_{\sigma_P} \cdot X\left(z,\psi_{g^{-1}}(u)\right))
\end{align*}
\end{corollary}

\begin{corollary}
If the action of $G$ on $P$ is transitive, defining $z \coloneqq p_{\sigma} = \sigma_P(p)$,  the dynamics of $[g]_{G_z} = \gamma_{\sigma_P}(p)$ evolve as
\begin{align*}
\dfrac{d}{dt} {[g]_{G_{z}}} =T_{[\id]_{G_{z}}}\bar{L}_{g_1} \left( T_{z}\gamma_{\sigma_P} \cdot X\left(z,\psi_{g_1^{-1}}(u)\right) \right)
\end{align*}
where $g_1 \in [g]_{G_z}$. 
\end{corollary}

\begin{assumption}
We will assume that either the map $H$ is $G$-equivariant or that $H$ is actually a map defined only on $\K \subset P$, that is, $H : \K \to \Y_2$. 
\end{assumption}
\begin{remark}
Note that we have  changed the definition of $H$ slightly, from what it was earlier in \eqref{eq:Output}, and the both can be reconciled by making minor technical changes. 
\end{remark}
If $H$ is restricted to $\K$ then we directly get a measurement involving only the base coordinate. Else if it is $G$-equivariant, then $H(g^{-1}\cdot p) = H(\sigma_P(_G[p])) = H(\oldtilde{\varphi}_g^{-1}(p))$ gives us a measurement of the base coordinate.

Combining the earlier results, we finally have the following reduction theorem:
\begin{theorem}
\label{thm:ReducedSystem}
The control system (\ref{eq:DynamicSystem}) -- (\ref{eq:Output}) is equivalent to the following system with $z \coloneqq \sigma_P(p)$, $[g]_{G_z} = \gamma_{\sigma_P}(p)$
\begin{align}
\dfrac{d}{dt} {[g]_{G_z}} &= T_{[\id]_{G_z}}\bar{L}_g \left( T_{z}\gamma_{\sigma_P} \left( X(z,\psi_{g^{-1}}(u)) - T_{_G[p]}\sigma_P(T_p\pi_{\phi} \cdot X(p,u)) \right) \right)
\label{eq:FibreDynamics}\\
y_{G} &=\varphi_{g^{-1}}(y_0)\label{eq:FibreOutput}\\
\dfrac{d}{dt}_G[p]&= \bar{X}(_G[p,u])\label{eq:BaseDynamics}\\
y_{\K}&=H(p)\label{eq:BaseOutput}
\end{align}
\end{theorem}

\begin{remark}
We therefore obtain a different way to express system (\ref{eq:DynamicSystem}) -- (\ref{eq:Output}) as system (\ref{eq:FibreDynamics}) -- (\ref{eq:BaseOutput}). The plan is to design an observer for this system by exploiting the symmetry highlighted by the new way of representation. Observe that if $H$ is restricted to $\K$ then the subsystem which evolves on the base manifold, that is (\ref{eq:BaseDynamics}) -- (\ref{eq:BaseOutput}), forms a subsystem independent of the fiber coordinate. Then we can design an observer for (\ref{eq:BaseDynamics}) -- (\ref{eq:BaseOutput}) first and use that to design an observer for (\ref{eq:FibreDynamics}) -- (\ref{eq:FibreOutput}). The observer for (\ref{eq:BaseDynamics}) -- (\ref{eq:BaseOutput}) is designed on a case by case basis, but for (\ref{eq:FibreDynamics}) -- (\ref{eq:FibreOutput}) we present a methodology in the next section.
\end{remark}

We present corollaries of Theorem \ref{thm:ReducedSystem} for the case when the group action is free or transitive. 

\begin{corollary}
If the action of $G$ on $P$ is free, the control system (\ref{eq:DynamicSystem}) -- (\ref{eq:Output}) is equivalent to the following system with $z \coloneqq \sigma_P(p)$, $g = \gamma_{\sigma_P}(p)$
\begin{align*}
\dfrac{d}{dt} {g}&= T_{\id}{L}_g ( T_{z}\gamma_{\sigma_P} \cdot X\left(z,\psi_{g^{-1}}(u)\right)) \\
y_{G} &=\varphi_{g^{-1}}(y_0) \\
\dfrac{d}{dt}_G[p]&= \bar{X}(_G[p,u]) \\
y_{\K}&=H(p)
\end{align*}
\end{corollary}

\begin{remark}
When the group action is free, \eqref{eq:FibreDynamics}--\eqref{eq:FibreOutput} becomes easier to handle since the fiber coordinate evolves on $G$ instead of $G/G_z$. 
\end{remark}

\begin{corollary}
If the action of $G$ on $P$ is transitive, the control system (\ref{eq:DynamicSystem}) -- (\ref{eq:Output}) is equivalent to the following system with $z \coloneqq p_{\sigma} = \sigma_P(p)$ and $[g]_{G_z} = \gamma_{\sigma_P}(p)$ 
\begin{align*}
\dfrac{d}{dt} {[g]_{G_z}} &=T_{[\id]_{G_{z}}}\bar{L}_{g_1} \left( T_{p_{\sigma}}\gamma_{\sigma_P} \cdot X\left(z,\psi_{g_1^{-1}}(u)\right) \right) \\
y_{G} &=\varphi_{g^{-1}}(y_0) 
\end{align*}
\end{corollary}

\begin{remark}
When the group action is transitive, choosing $p_{\sigma}$, makes \eqref{eq:BaseDynamics}--\eqref{eq:BaseOutput} moot and reduces the problem to that of designing an observer on the fibres (i.e. the Lie group), that is, designing a filter for (\ref{eq:FibreDynamics}) -- (\ref{eq:FibreOutput}). This physically corresponds to the case of the unicycle and rigid body attitude observation using IMUs considered in \cite{mahony-2013} and SLAM with known spatial markers that was considered in \cite{mahony-2017}.
\end{remark}

\section{Gradient Based Observers for Kinematic Systems on Lie Groups}
\label{sec:liegroupobs}
In this section we present observer design when the system evolves on a Lie group. This is a restatement of results in \cite{mahony-2013}, but in our setting, presented with the intention of cementing the ideas that we discuss. 
This corresponds to designing an observer for (\ref{eq:FibreDynamics}) -- (\ref{eq:FibreOutput}). 
This physically corresponds to the situations dealt with in \cite{mahony-2013},\cite{mahony-2017}.

Let $\Phi :G \times G \to G$ be group multiplication. We will assume $\Phi$ is a left action, that os, $\Phi(g_1,g_2) = \Phi_{g_1}(g_2) = g_1g_2$ 
and let $\varphi : G\times \mathcal{Y}\to \mathcal{Y}$ be a proper and free $\Phi$-invariant left group action of $G$ on $\mathcal{Y}$, that is  $\varphi(h,\varphi(g,y))=\varphi\left(\Phi_h(g),y\right)$. 
Let  $\left\langle\left\langle \cdot,\cdot\right\rangle \right\rangle: \g \times \g \to \R$ be an inner product on the lie algebra.

We consider a system that evolves according to
\begin{align}
\dot{g}&=T_{\id}\Phi_g \cdot \zeta(t),\label{eq:Kineamtics}\\
y&=\varphi_{g^{-1}}(y_0),\label{eq:OutputKinematic}
\end{align}
where $\zeta(t)\in \mathcal{G}$ is known and $y_0\in \mathcal{Y}$ is a constant. {This corresponds to (\ref{eq:FibreDynamics}) -- (\ref{eq:FibreOutput}) where we have treated $\sigma_P([p](t))$ as an input.

The problem we consider is that of estimating $g$ from the measurement of $y$ given the information of $\zeta$.
We consider the pre-observer
\begin{align}
\dot{\widetilde{g}}&=T_{\id}\Phi_{\tilde{g}}\cdot (\zeta-\Delta(\widetilde{g},{y})), \label{eq:pre-obs-group}\\
\widetilde{y}&=\varphi_{\widetilde{g}^{-1}}(y_0). \notag
\end{align}
$\tilde{g}$ is termed the estimate, and $\Delta$ is termed the innovation. The innovation will be designed now so that $\tilde{g}$ converges to $g$. 
Note that the innovation is a Lie-algebraic valued function of the estimate and the measurement.

Consider $V^y(\cdot,\cdot): \mathcal{Y}\times \mathcal{Y} \to \mathbb{R}$ such that $V^y(\varphi_{g}({y_1}),\varphi_{g}({y_2}))=V^y(y_1,y_2)$ for all $g\in G$ and ${y_1},{y_2}\in \mathcal{Y}$.  Also, $V^y(y_1,y_2) = V^y(y_2,y_1)$ for all $y_1,y_2 \in \Y$. This is called a distance function. 

Consider the estimation error and the output error given respectively by
\begin{align*}
e_g&\triangleq \Phi_g(\widetilde{g}^{-1}),\\
e_y&\triangleq V^y(\varphi_{g^{-1}}(y_0),\varphi_{\tilde{g}^{-1}}(y_0))=V^y(\varphi_{e_g}(y_0),y_0) \eqqcolon V^e(e_g).
\end{align*}

\begin{theorem}
Suppose that $\id$ is a non-degenerate critical point for $V^e$.
Define $\zeta_e \in \mathcal{G}$ as  
\begin{align}
\left\langle (\Phi_{e_g})^*d_{e_g}V^e\:,\:\cdot \right\rangle
=\left\langle\left\langle \zeta_e\:,\:\cdot\right\rangle \right\rangle
\label{eq:zeta-e-defn}
\end{align}
where $d_{e_g}V^e$ is the differential of $V$ at $e_g$ and $(\Phi_{e_g})^*$ denotes the pullback. If
\begin{align*}
\Delta(\widetilde{g},y)&\triangleq 
-k\mathrm{Ad}_{\tilde{g}^{-1}}\zeta_e 
\end{align*}
then there exists a nighbourhood $\mathcal{B}$ of $\id$ such that if $e_g \in \mathcal{B}$ then $e_g$ converges to $\id$. 
\end{theorem}
\begin{proof}

The first step in the proof is to show that $\zeta_e$ is independent of $g$. This is essential since we use $\zeta_e$ in the construction of the observer.

\begin{claim}
$\zeta_e$ is independent of $g$.
\end{claim}
\begin{proof}
We know that
\begin{align*}
 V^y(\varphi_{g^{-1}}(y_0),\varphi_{\tilde{g}^{-1}}(y_0))&=V^y(y_0,\varphi_{g\tilde{g}^{-1}}(y_0))=V^y(y_0,\varphi_{e_g}(y_0))= V^e(e_g) 
\end{align*}

For an arbitrary $\xi \in \g$, consider the curve $c(s)={e_g}\exp{(\xi s)}$ that passes through $e_g$ at $s=0$ with tangent vector $T_IL_{e_g}\cdot \xi$. Let $\langle \cdot , \cdot \rangle$ denote the duality pairing between a covector and vector. Then  \eqref{eq:zeta-e-defn} can be written as
\begin{align*}
\left\langle\left\langle \zeta_e\:,\:\xi\right\rangle \right\rangle = &\langle  (T_{\id}L_{e_g})^*d_{e_g}V^e\,,\,\xi\rangle=\langle d_{e_g}V^e\,,\,e_g\cdot\xi\rangle=\left.\dfrac{d}{ds}\right|_{s=0}V^e({e_g}\exp{(\xi s)}) \\ =&\left.\dfrac{d}{ds}\right|_{s=0}V^y(\varphi_{e_g\exp{(\xi s)}}(y_0),y_0),
=\left.\dfrac{d}{ds}\right|_{s=0}V^y(\varphi_{g\widetilde{g}^{-1}\exp{(\xi s)}}(y_0),y_0) \\ =& \left.\dfrac{d}{ds}\right|_{s=0}V^y(\varphi_{\widetilde{g}^{-1}\exp{(\xi s)}}(y_0),y).
\end{align*}

The right hand side depends only on $\tilde{g}$, $y$ and $\tilde{y}$ for any $\xi$ thus $\zeta_e$ depends only on $\tilde{g}$, $y$ and $\tilde{y}$.
\end{proof}

From the definitions of $e_g$ and $V^e$ we have
\begin{align*}
\dot{e}_g&=\Phi^{\tilde{g}}_{e_g} \Delta(\widetilde{g},y)\\
\dot{V}^e&=\left\langle d_{e_g}V^e\,,\, \Phi^{\tilde{g}}_{e_g}\cdot\Delta(\tilde{g},y)\right\rangle
\end{align*}
where
\begin{align*}
\Phi^{\tilde{g}}_{e_g}&\triangleq 
T_{\id}\Phi_{e_g}\circ\mathrm{Ad}_{\tilde{g}} 
\end{align*}

It is now easy to see that the innovation term
\begin{align*}
\Delta(\widetilde{g},y)&\triangleq 
-k\mathrm{Ad}_{\tilde{g}^{-1}}\zeta_e 
\end{align*}
will yield the time invariant error dynamics
\begin{align}
\dot{e}_g&=T_{\id}\Phi_{e_g}\left(-k\zeta_e\right), \nonumber\\
\dot{V}^e&=-k\,\langle\langle\zeta_e\,,\,\zeta_e\rangle\rangle \le 0 \label{eq:Vdot-le-zero}
\end{align}

From the definition of $V^e$ it is clear that $V^e(\id) = 0$. Further since we assume that $\id$ is a non-degenerate critical point for $V^e$, then $\id$ is an isolated critical point \cite[Corollary 2.3]{milnor}. Thus there exists a nighbourhood $\mathcal{B}$ of $\id$ such that if $e_g \in \mathcal{B}$ then the condition $\dot{V}^e \le 0$ in \eqref{eq:Vdot-le-zero} ensures that $e_g$ converges to $\id$. 
\end{proof}

Let us conclude this section with an example. 

\begin{example}
\label{ex:LG-obs}
Consider the case of attitude kinematics with $G = \SOthree$ measured body angular velocities, $\hat{\Omega} \in \sothree$, where the kinematic equations are 
\begin{align*}
\dot{R}&=R\hat{\Omega}
\end{align*}
where we let $\hat{(\cdot)} : \R^3 \ra \mathfrak{so(3)}$ be the canonical isomorphism between the two. The measured outputs are two non-collinear inertial directions $e_3$ and $e_2$ in the body frame. $e_2$ and $e_3$ are assumed to be known. $\mathcal{Y}=\mathbb{S}^2\times \mathbb{S}^2\subset \mathbb{R}^3\times \mathbb{R}^3$ and $y=(y_2,y_3)=(R^Te_2,R^Te_3) =(\varphi_{R^T}(e_2),\varphi_{R^T}(e_3))$ where $\varphi \,: \,\SOthree \times \mathbb{S}^2 \to \mathbb{S}^2$ is a left action that is simply given by multiplication by $R$. Table \ref{tb:attobssummary} contains a summary of the structure. 

\begin{table}[h!]
\centering
 \begin{tabular}{|c|c|c|} 
 \hline
  $P =\R^3 \setminus \{0\}$ & $G = \SOthree$ & $\mathcal{Y} = \mathbb{S}^2\times \mathbb{S}^2$\\ \hline 
  \end{tabular}
 \caption{Summary of Structure}
 \label{tb:attobssummary}
\end{table}
Let the pre-observer, as per \eqref{eq:pre-obs-group}, be
\begin{align}
\dot{R} = R(\hat{\Omega} - \Delta)
\label{eq:pre-obs-SO3}
\end{align}
To complete the observer design, choose a cost function
\begin{align*}
V^y(y,\tilde{y})&\triangleq ||R^T\,e_3-\tilde{R}^T\,e_3||^2+||R^T\,e_2-\tilde{R}^T\,e_2||^2 \\
&=||e_3-R\tilde{R}^T\,e_3||^2+||e_2-R\tilde{R}^T\,e_2||^2,\\
&= ||e_3-E\,e_3||^2+||e_2-E\,e_2||^2 \eqqcolon V^e(E)
\end{align*}
Let $c(s)=E\exp{(\widehat{\xi}s)}$
\begin{align*}
\langle (TL_{E})^*d_{E}V^e\,,\,\xi\rangle&=\langle d_{E}V^e\,,\,E\widehat{\xi}\,\,\rangle=\dfrac{d}{ds}|_{s=0}V^e({E}\exp{(\widehat{\xi} s)})\\
&=\sum_{k=2}^3\left.\dfrac{d}{ds}\right|_{s=0}||e_k-{E\exp{(\widehat{\xi} s)}}e_k||^2,\\
&=\sum_{k=2}^3\left.\dfrac{d}{ds}\right|_{s=0}||R^Te_k-{\tilde{R}^T\exp{(\widehat{\xi} s)}}e_k||^2 \\
&=\sum_{k=2}^3\left.\dfrac{d}{ds}\right|_{s=0}||y_k-{\tilde{R}^T\exp{(\widehat{\xi} s)}}e_k||^2\\
&=\sum_{k=2}^3\left.\dfrac{d}{ds}\right|_{s=0}\left(y_k^Ty_k-2y_k^T{\tilde{R}^T\exp{(\widehat{\xi} s)}}e_k+e_k^Te_k\right) \\
&=\sum_{k=2}^3-2y_k^T{\tilde{R}^T\widehat{\xi}}e_k =\sum_{k=2}^3\tr(-2e_ky_k^T\tilde{R}^T\hat{\xi})\\ 
&= \sum_{k=2}^3\tr\left(\left(-2\tilde{R}y_ke_k^T\right)^T\hat{\xi}\right)
\end{align*}
If the iner product on $\g$ is the Frobenius inner product, we get (after considering the skew symmetric part of $-2\hat{e}_k\tilde{R}y_k$ in order to get an element in $\g$),
\begin{align*}
 \langle (TL_{E})^*d_{E}V^e\,,\,\xi\rangle
= \sum_{k=2}^3\tr\left(\left(-\tilde{R}y_ke_k^T + e_ky_k^T\tilde{R}^T\right)\hat{\xi} \right)
\end{align*}
which yields 
\begin{align}
\zeta_e =& \sum_{k=2}^3 \left(-\tilde{R}y_ke_k^T + e_ky_k^T\tilde{R}^T \right) \notag \\
 \Delta
= & -k\sum_{k=2}^3\left( -y_ke_k^T\tilde{R} + \tilde{R}^Te_ky_k^T \right) \label{eq:innovation-SO3}
\end{align}
Substitute \eqref{eq:innovation-SO3} into the pre-observer \eqref{eq:pre-obs-SO3} to get the final observer structure.
\end{example}

\section{Examples}
\label{sec:examples}

In this section we present two examples to illustrate some of the concepts developed so far. We continue example \ref{eq:SOonR3} and show its relevance to a target tracking problem. In the second example, we show how the SLAM problem falls into our framework. It helps us emphasise the case of the free group action.

\subsection{Target tracking problem}
In this example, we consider a point object whose trajectory $p(\cdot)$ evolves on $\R^3 \setminus \{0\}$. We will first look at how the kinematics decompose into two subsystems as per the base and fiber spaces, and then look at giving a geometric interpretation to the well known problem of target tracking using range and bearing measurements.

}

\subsubsection{Decomposition of a kinematic system}

We continue example \ref{eq:SOonR3} to analyse the decomposition of kinematics. Table \ref{tb:spheresummary} provides a summary of the structure in the problem.

\begin{table}[h!]
\centering
 \begin{tabular}{|c|c|c|} 
 \hline
  $P =\R^3 \setminus \{0\}$ & $G = \SOthree$ & $\phi(g,p) = gp$  \\
  \hline  
 \end{tabular}
 \caption{Summary of Structure}
 \label{tb:spheresummary}
\end{table}

Let us now look at how the kinematics of its motion split as per the base and fibre co-ordinate structure. Define $\R \ni t \mapsto r(t) := ||p(t)||_2 \in \R$. Let $R(\cdot)$ be a smooth curve in $\SOthree$ be such that $p(t) = r(t)R(t)p_0$ (using Given's rotations \cite{watkins}). 
Let $\R \ni t \mapsto v(t) \coloneqq \dot{p}(t) \in \R$. Assume further that $\dot{R}(t) = R{\hat{\Omega(t)}}$ for some $\Omega(t) \in \R^3$ and recall that $\hat{(\cdot)} : \R^3 \ra \mathfrak{so(3)}$ is the canonical isomorphism between the two. We shall suppress the explicit time argument henceforth to keep the notation terse.

\begin{align*}
\dot{p} &= v \\
&= \dot{r}Rp_0 + r\dot{R}p_0 \\
&= \dot{r}Rp_0 + rR{\Omega}_{\times}p_0 
\end{align*}

Recall from example \ref{eq:SOonR3}, that the base co-ordinate of $p$ is $rp_0$ and the fibre co-ordinate is $[R(t)]_{G_{p_0}}$. Recall further that 
\begin{align*}
\ver_p(v) &= \dot{r}Rp_0 + rR{\Omega}_{\times}p_0 - r(Rp_0)^T(\dot{r}Rp_0 + rR{\Omega}_{\times}p_0)\frac{rRp_0}{r^2} \\
&= rR{\Omega}_{\times}p_0 \\
\hor_p(v) &=  \dot{r}Rp_0
\end{align*}

Therefore, we see how the kinematics splits into two smaller subsystems. The evolution of the horizontal component depends only on the base coordinate, therefore it becomes an independent subsystem in itself. Once a solution to the horizontal subsyetm is obtained, one can proceed to solving the vertical subsystem. We now show how given a range-bearing measurement model, one can infer the base and fiber coordinate and thus reconstruct the total state.

\subsubsection{Application given range and bearing measurements}
\begin{figure}
\centering
\includegraphics[scale=0.8]{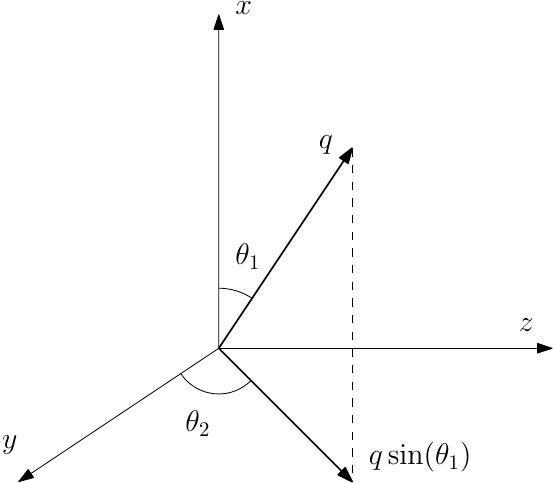}
\caption{Radar}
\label{fig:radar}
\end{figure}

Suppose that there is a target tracking problem in which the user is interested in tracking the particle whose position is given by $p(\cdot)$ as a function of time, and measurements are available to use in the form of range and bearing
\begin{align*}
y_1 &= r \\
y_2 &= \theta_1 \\
y_3 &= \theta_2
\end{align*}

where $\theta_1$ and $\theta_2$ are marked in Figure \ref{fig:radar} (they are respectively the polar and azimuthal angles in spherical polar co-ordinates). This is a typical range and bearing measurement model, used in various problems like submarine target tracking, ground target tracking to name a few applications, and readers are redirected to \cite{bishop-2007,ristic,mallick,mallick-2015} for more on this. Define 
\begin{align*}
R_0 \coloneqq \begin{pmatrix} \cos(\theta_1) & -\sin(\theta_1) & 0 \\ \sin(\theta_1)\cos(\theta_2) & \cos(\theta_1)\cos(\theta_2) & -\sin(\theta_2) \\ \sin(\theta_1)\sin(\theta_2) & \cos(\theta_1)\cos(\theta_2) & cos(\theta_2) \end{pmatrix}
\end{align*}

Then the base co-ordinate is directly given by the first measurement $y_1$ and the fibre co-ordinate is given by $[R_0]_{G_{p_0}}$. This gives us a geometric interpretation of the well known range bearing measurement problem.

\subsection{SLAM}
\subsubsection{Modelling}
We recall the mathematical modelling of the SLAM problem from \cite{mahony-2017}. 
Assume that there is a fixed inertial frame represented by $\I$ and there is a body frame represented by $\B$. Assume that there are $N$ fixed but unknown landmarks represented by $\L_i \in \R^3$ for $i = 1,2,\ldots N$ in $\I$. The $\L_i$ are measured in $\B$ as $l_i \in \R^3$. Assume that $S = \begin{pmatrix}
R & r \\ 0 & 1 \end{pmatrix} \in \SEthree$ represents the Euclidean transformation between $\B$ and $\I$. Then $Rl_i + r= \L_i$. The goal is to observe $S$ and all $\L_i$. 

To easily represent the action of $\SEthree$ on $\R^3$, we will define
\begin{align*}
&\E^3 \coloneqq \set{\begin{pmatrix} l \\ 1 \end{pmatrix}}{l \in \R^3}, \quad \bar{\R}^3 \coloneqq \set{\begin{pmatrix} l \\ 0 \end{pmatrix}}{l \in \R^3} \\
&\E \coloneqq \overbrace{\E^3 \times \E^3 \times \cdots \times \E^3}^{N \text{ times}}, \quad \euc{\R} \coloneqq \overbrace{\euc{\R}^3 \times \euc{\R}^3 \times \cdots \times \euc{\R}^3}^{N \text{ times}} 
\end{align*}

Each $l_i$ (or $\L_i$) will be denoted as elements of $\E^3$ as 
\begin{align*}
\euc{l}_i \coloneqq \begin{pmatrix} l_i \\ 1 \end{pmatrix}, \quad \euc{\L}_i \coloneqq \begin{pmatrix} \L_i \\ 1 \end{pmatrix} \implies S\euc{l}_i = \euc{\L}_i
\end{align*}

The state space is $P = \SEthree \times \E$. 
Hence $p \coloneqq (S , \lb_1 , \lb_1 , \ldots , \lb_N ) \in P$.
The Lie group is $G = \SEthree$ and the group action $\phi$ (a free right action) is
\begin{align*} 
\SEthree \times P \ni (g,p) \mapsto \phi(g,p):= (g^{-1}S, g^{-1}\lb_1, g^{-1}\lb_2, \ldots , g^{-1}\lb_N ) \in P
\end{align*} 
The measurement is 
\begin{align}
P \ni p \mapsto H(p) \coloneqq (\l_1 , \l_2 , \ldots , \l_N ) = ( S^{-1}\lb_1 , S^{-1}\lb_2 , \ldots , S^{-1}\lb_N ) \in \E
\label{eq:SLAM-msm}
\end{align}

The input space $\mathcal{U}$ is $\mathfrak{se(3)} \times \bar{\R}$. Define the action of the group on the input space, $ G \times \mathcal{U} \ni (g,v) \mapsto \psi(g,v) \coloneqq v \in \mathcal{U}$.
The kinematics of SLAM are 
\begin{align}
\dot{p} = X(p,v) \text{ with } X(p,v) \coloneqq (SV,S\v_1,...,S\v_N) \text{ and } v \coloneqq (V,\v_1,\ldots,\v_N) \in \mathcal{U}
\label{eq:SLAM-kin}
\end{align}
 where $v$ is termed the velocity and is assumed to be measured. It is easy to verify that $T_p\phi_g X(p,v) = X(\phi_g(p),\psi_g(v))$. The structure is tabulated in Table \ref{tb:slamsummary}.

\begin{table}[h!]
\centering
 \begin{tabular}{|c|c|c|} 
 \hline
  $P =\SEthree \times \E$ & $G = \SEthree$ & $\Y = \E$\\ \hline 
  \end{tabular}
 \caption{Summary of Structure}
 \label{tb:slamsummary}
\end{table}

\subsubsection{Geometry}
Let us now investigate the geometry behind the modelling of the problem. 
The orbit of $p \in P$ is 
\begin{align*}
\mathcal{O} (p) = \set{(g^{-1}S, g^{-1}\lb_1, g^{-1}\lb_2, \ldots, g^{-1}\lb_N)}{g \in \SEthree}
\end{align*} 
The quotient space space thus formed is $P/G = \set{_G[p]}{p \in \T}$. Define $P \ni p \mapsto \pi(p) \coloneqq _G[p] \in P/G$.

Choose a $W \in \mathfrak{se}(3)$, then
\begin{align*}
T_I\phi^p \cdot W = (-WS,-W\lb_1,\ldots, -W\lb_N).
\end{align*}
Thus $T_p\O(p) = \set{T_I\phi^p \cdot W}{W \in \mathfrak{se}(3)}$. 
	Consider a smooth curve $\beta(t) := \phi^p(g(t))$ where $g(\cdot)$ is a smooth curve on $\SEthree$ with $g(0) = I$, $\dot{g}(0) = W$. Observe that $\dot{\beta}(0) = (-WS,-W\lb_1,\ldots, -W\lb_N)$ Since $\pi(\beta(t)) = _G[p]$ then $\dot{\beta}(0) \in \ker T_p\pi$. 
	Hence 
\begin{align*}	
	\Ver(p) = \set{(-WS,-W\lb_1,\ldots,-W\lb_N)}{W \in \mathfrak{se}(3)}
\end{align*} 
It can be seen that $\Ver(p) = T_p\O(p)$.	
	One choice for horizontal space is
	\begin{align*}	
	\Hor(p) = \set{(0,\v_1,\ldots,\v_N)}{\v_1,\ldots,\v_N \in \bar{\R}^3}
	\end{align*} 
	
	Recall that, to confirm whether this choice is indeed a valid choice for a horizontal space, we need to verify if $\Hor(g \cdot p) = \Hor(p)$ and $\Hor(p) \cap \Ver(p) = \{0\}$. Since $\Hor(p)$ is the same set for all $p$, $\Hor(g \cdot p) = \Hor(p)$. Now let $v = \in \Hor(p) \cap \Ver(p)$. Then proving $v = (0,\v_1,\ldots,\v_N)$ for some $\v_1,\ldots,\v_N \in \bar{\R}^3$ and $v = (-WS,-W\lb_1,\ldots,-W\lb_N)$ for some $W \in \mathfrak{se}(3)$. Therefore, $WS = 0$ which implies $W = 0$ since $S$ is invertible. Hence $v = 0$ and $\Hor(p) \cap \Ver(p) = \{0\}$. Since $\dim(\Ver(p)) + \dim(\Hor(p)) = \dim(T_p\T)$, $\Ver(p) \bigoplus \Hor(p) = T_p\T$ for this choice of $\Hor(p)$. Given any $v_p = (SV, \v_1,\v_2,\ldots,\v_N) \in \T$, define $W \coloneqq -SVS^{-1}$ then
\begin{align*}	
	\ver(v_p) &= (-WS,-W\lb_1,-W\lb_2,\ldots,-W\lb_N) \\
	 \hor(v_p) &= (0, \v_1 +W\lb_1,\v_2 +W\lb_2,\ldots,\v_N +W\lb_N) 
\end{align*}
	
	Consider a choice of section, $P/G \ni _G[p] \mapsto \sigma_P(_G[p]) := (\id,S^{-1}\lb_1,\ldots,S^{-1}\lb_n) \in P$ and it is easy to observe that $\pi \circ \sigma = id_{P/G}$. Defining $\T \ni p \mapsto \gamma_{\sigma_P}(p) := S^{-1} \in \SEthree$, we see that $p = \phi(\gamma_{\sigma_P}(p),\sigma_P(_G[p]))$. Define $P \supset \K \coloneqq \sigma(P/G) = \{\id\} \times \E$. It is evident that $T_z\K = \Hor(z)$ for all $z \in \K$, implying that $\K$ intersects each orbit transversally. Therefore, we can identify any $p \in P$ with 
\begin{itemize}	
	\item $S \in G$ (or equivalently, $S^{-1} \in G$), which is the fibre coordinate and 
	\item $z \coloneqq (\id,S^{-1}\lb_1,\ldots,S^{-1}\lb_n) \in \K$, which is the base coordinate.
\end{itemize}	 
$\K$ is representative of the base manifold, and therefore, SLAM admits a global cross-section, which greatly simplifies the process of representing a point in base and fibre coordinates.

\subsubsection{Kinematics and Observer Design}
The SLAM kinematics and measurement in 	\eqref{eq:SLAM-kin}, \eqref{eq:SLAM-msm} form a control system, which decomposes as as follows in light of the geometric structure highlighted above
\begin{align*}
	&\dot{z} = (0,-VS^{-1}\lb_1 + \v_1,\ldots,-VS^{-1}\lb_N + \v_N) \\
	&y_{\K} = H(p) \\
	&\frac{d}{dt}(S^{-1})= -VS^{-1}
	\end{align*}

The kinematics of the fiber coordinate $S$ can equivalently be written as $\dot{S} = SV$. The dynamics of the base coordinate $z$ are governed by the horizontal components of the kinematic vector field while that of the fiber coordinate are governed by the vertical component. The problem therefore decomposes into two smaller independent subsystems. 

The measurement $y_{\K}$ directly yields the base coordinate since $(\id, H(p)) = z$. There is however a lack of a measurement of the fiber coordinate. This can be explained by the fact that SLAM problem thus modelled, is not fully observable; it can only be observed upto a Euclidean transformation \cite{wang-2018,mahony-2017,kwlee-2018}. 
%
%
%
%

%
To fill this gap, we will introduce new measurements 
to construct an observer for the fiber coordinate (although these measurements have been given a physical justification, they are not available to the user in conventional applications). However, since our paper focusses on exploiting geometric structure in the design of observers, we go ahead with modifying the traditionally known SLAM problem, so that we can design an observer for the fiber coordinate.
To this end, introduce known and fixed $\mathscr{L}_i \in \R^3$ for $i = 1,2,\ldots,\mathscr{N}$ and $\bar{\mathscr{L}}_i \coloneqq \begin{pmatrix} \mathscr{L}_i \\ 1 \end{pmatrix}$ ($\mathscr{L}_i$ is known for all $i$ as opposed to $\mathcal{L}_i$ which was unknown). Define the measurement 
\begin{align*}
y_G \coloneqq (S^{-1}\bar{\mathscr{L}}_1,S^{-1}\bar{\mathscr{L}}_2,\ldots,S^{-1}\bar{\mathscr{L}}_{\mathscr{N}})
\end{align*}
Suppose that $\mathscr{L}_i$ for $i = 1,2,\ldots,\mathscr{N}$ satisfy Assumption 1 in \cite{mahony-2011}, which is a technical assumption similar to non-collinearity in the Example \ref{ex:LG-obs} useful while proving observer convergence. Then, utilising the measurement $y_{G}$, an observer can be designed for the fiber coordinate $S$ as per the methodology in \cite{mahony-2011}. 


\section*{Acknowledgments} The authors would like to thank Debasish Chaterjee, Navin Khaneja, Amit Sanyal and Srikant Sukumar for helpful discussions. A part of the third author's work was supported by a MATRICS fellowship from the Department of Science and Technology.

\providecommand{\href}[2]{#2}
\providecommand{\arxiv}[1]{\href{http://arxiv.org/abs/#1}{arXiv:#1}}
\providecommand{\url}[1]{\texttt{#1}}
\providecommand{\urlprefix}{URL }

\medskip
Received xxxx 20xx; revised xxxx 20xx.
\medskip

\end{document}